# Hysteresis of Electronic Transport in Graphene Transistors

Haomin Wang, [a), *] Yihong Wu

Department of Electrical and Computer Engineering, National University of Singapore, 4 Engineering Drive 3, Singapore 117576

Chunxiao Cong, Jingzhi Shang, and Ting Yu [a)]

Division of Physics and Applied Physics, School of Physical and Mathematical Sciences, Nanyang Technological University, 21 Nanyang Link, Singapore 637371

a) Electronic mail: haomin.wang@gmail.com; yuting@ntu.edu.sg

* Present address: Division of Physics and Applied Physics, SPMS, Nanyang Technological University, Singapore 637371

**ABSTRACT** Graphene field effect transistors commonly comprise graphene flakes lying on $SiO_2$ surfaces. The gate-voltage dependent conductance shows hysteresis depending on the gate sweeping rate/range. It is shown here that the transistors exhibit two different kinds of hysteresis in their electrical characteristics. Charge transfer causes a positive shift in the gate voltage of the minimum conductance, while capacitive gating can cause the negative shift of conductance with respect to gate voltage. The positive hysteretic phenomena decay with an increase of the number of layers in graphene flakes. Self-heating in helium atmosphere significantly removes adsorbates and reduces positive hysteresis. We also observed negative hysteresis in graphene devices at low temperature. It is also found that an ice layer on/under graphene has much stronger dipole moment than a water layer does. Mobile ions in the electrolyte gate and a polarity switch in the ferroelectric gate could also cause negative hysteresis in graphene transistors. These findings improved our understanding of the electrical response of graphene to its surroundings. The unique sensitivity to environment and related phenomena in graphene deserve further studies on nonvolatile memory, electrostatic detection and chemically driven applications.

**KEYWORDS:** graphene transistor, conductance hysteresis, charge transfer, capacitive gating, water dipole

The recent discovery of graphene [1] filled the gap between one-dimensional carbon nanotubes and three-dimensional graphite in graphitic materials, according to dimensional classification. It makes truly two dimensional crystals accessible and links solid state devices to molecular electronics.[2, 3] Graphene is identified as a zero gap semiconductor and has a novel electronic structure with its conduction band and valence band touching each other at a neutral point. This characteristic enables graphene devices to exhibit ambipolar behavior, despite having an unexpectedly high conductivity minimum. This observation was recently explained by the formation of electron hole puddles at the Dirac point,[4] which arise from charged impurity centers [5-7] or graphene ripples [8]. The nature of the low density of states near the Dirac point makes the electronic properties of graphene very sensitive to the surroundings. In addition, large surface-to-volume ratio allows graphene to catch many adsorbates very easily. Recent transmission electronic microscopy (TEM) experiment clearly reveals the existence of adsorbates on the surface of graphene. [9, 10] The transport properties across and parallel to the interface of graphene may be changed by the manner of sticking and bonding the atoms/molecules. [11, 12] In these ways, the environment has a direct impact in the physical properties of graphene. Hence, understanding the graphene-environment interaction is fundamental in either interpreting graphene sensing phenomena or engineering graphene sheets for functioning and high performance devices.

In many studies, graphene field effect transistors commonly comprise graphene flakes lying on an insulator and back-gated by a silicon substrate. Hysteretic behaviors in conductance characteristics with gate are often observed. [13-22] The hysteresis varies depending on the sweeping voltage range, sweeping rate, and surrounding conditions. The hysteresis presumably originates from charge transfer from neighboring adsorbates (such as: water molecule) or charge injection into the trap sites on dielectric substrate in general. Vacuum treatment and annealing were found to effectively suppress the hysteresis.[13, 14] In addition, a hydrophobic polymer layer was introduced in between graphene and $SiO_2$ to suppress the hysteresis. [19, 20] On the other hand, as high-k media, water/ice layer could greatly enhance the transport performance of graphene FET by dielectric screening and may suppress hysteresis.[23, 24] There appears to be no broad consensus on mechanisms that are tempting to attribute the hysteresis. In summary, although interesting data have been presented in these recent reports, several important physical questions have yet been addressed in detail: (1) Is the origin of hysteresis the same in graphene and carbon nanotube transistors? (2) What are the mechanisms governing the hysteresis in the electrical characteristics? 3) How long is the relaxation time for the hysteresis? 4) What is the role of adsorbed water molecule on/below graphene in hysteresis? In fact, the hysteresis may cause uncertainty in measuring both conductance and mobility, which may lead to large discrepancies in the results once any sweeping rate/range/direction from the gate was not reported. Therefore, it is worth systematically investigating the hysteresis and understanding its origin. The investigation will benefit research on high performance graphene transistors, graphene sensors and nonvolatile memory electronics.

In this work, we present a systematic investigation of the conductance *via* sweeping gate bias by forward and backward scans in graphene transistors with dielectric and electrolyte gating. First, we study the transport properties of graphene field effect transistors on $SiO_2$, and address the scaling of hysteresis with respect to the gate voltage sweeping range, sweeping rate, and the number of graphene layers. As the parallel electrical field between graphene and back gate is greatly different than the radiating one near carbon nanotube, which could cause the local breakdown of $SiO_2$, the chance for charge injection into the dielectric substrate is much lower than that of charge trapping into neighboring dipolar adsorbates (such as, water molecules) in graphene transistor. Following that, we examine the hysteresis in electrolyte-gated graphene transistor. Dramatic hysteresis of conductance is different from the one in back-gated graphene transistor. What is more, we find that back-gated graphene transistors exhibit the two kinds of different hysteresis by sweeping the gate bias above and below 0°C. The result indicates that ice on/below graphene exhibits stronger dipole moment than water does. The action of hysteresis is believed to be attributed to either charge trapping from/to graphene or capacitive gating effect. The two mechanisms generally coexist and compete with each other in all graphene transistors.

**RESULTS AND DISCUSSION**

We began by measuring the hysteresis of graphene field effect transistors on $SiO_2$, which were fabricated by method 1. (See Methods or ref. [25, 26]) For the sake of comparison, all of the investigated graphene devices presented reproducible hysteresis in conductance versus gate potential ($V_{back-gated}$ which is rewritten to $V_{bg}$) characteristics in ambient environment. The gate voltage was swept continuously from -80 V to 0 V, then to +80 V and back to -80 V. A representative example is displayed in Figure 1. Hall bar was adapted as the electrode geometry. As shown in Figure 1(b), the gate scans lead to the positive shift of the neutral point (NP) in the back gated system, where we define the NP is the voltage at conductance minimum. We could determine the direction of the hysteresis by comparing the location and the sign of trapped charges to the avalanching field. When the back gate starts at negative gate voltage, holes in graphene are slowly trapped into the trap centers, so that after some time the graphene sees a more positive potential than that simply due to the gate voltage (and vice versa for the opposite sweep direction). These trapped charges under graphene dope the graphene into opposite polarity, whose static condition could be observed in scanning single electron transistor [4]. Charge traps seem to be charged on time scales comparable to the scale relevant for the measurement. As such, gate voltage sweeping in negative regime typically shifts the NP down because of charge screening from injected holes into trap sites. Similarly, gate voltage sweeping in positive regime induces electron injection into the trap sites, and NP is shifted up, In DC gate sweeping, the charges remain trapped until the gate polarity is switched.

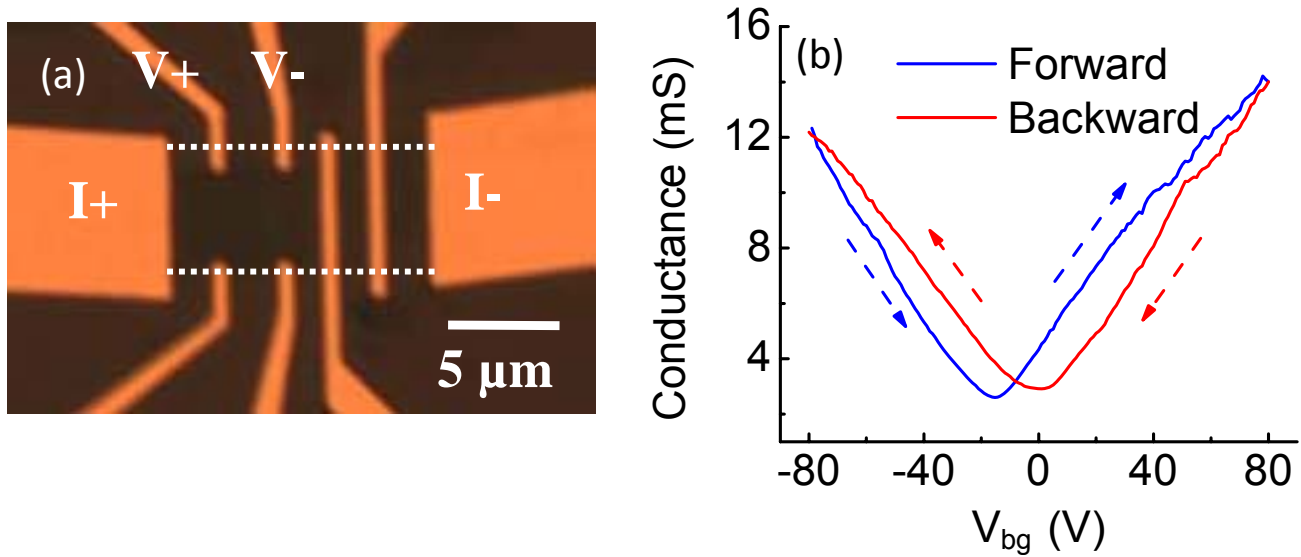

FIGURE 1. (Color online) a) Optical image of a bilayer graphene device (bs4q3p7) lying on $SiO_2$; b) Conductance vs gate voltage curves recorded under sweep rate of 1.25 V/s in ambient condition. As the gate voltage is swept from negative to positive and back, a pronounced hysteresis is observed, as indicated by the arrows denoting the sweeping direction.

As shown in Figure 2 (a), the magnitude of the hysteresis in $V_{bg}$ exhibits a significant dependence on sweeping rate of the gate voltage ($dV_{bg}/dt$). The voltage difference of NP caused by hysteresis increases from 8 V ($dV_{bg}/dt$ = 5 V/s) to 14.5 V ($dV_{bg}/dt$ = 1.25 V/s) and further to 15 V $dV_{bg}/dt$ = 0.313 V/s). Device hysteresis increases steadily with reducing the sweeping rate because of charge trapping on a time scale longer than few seconds. It almost saturates when the sweeping rate is lower than 1.25 V/s, and therefore, the sweeping rates are adjusted in subsequent experiments unless otherwise noted. Figure 2 panels (b) &(c) compare conductance ~ $V_{bg}$ curves of the graphene FET for different voltage ranges at room temperature. The hysteresis becomes larger as the range of $V_{bg}$ increases. It indicates that hysteresis originates from charge injection into interface or bulk oxide traps at higher gate bias. When the voltage is in the range of 10 V, the NP shift due to hysteresis is only about 1 V. When the range of $V_{bg}$ increases beyond 10 V, the hysteresis becomes more pronounced. The NP shift goes up to about 4.3V (in the range of 40 V) and 15 V (in the range of 80 V). The enhanced hysteresis can be quantitatively explained here. Silicon dioxide has interface and bulk (dangling bond) traps whose charge state changes with gate voltage. At the $Si/SiO_2$ interface of conventional silicon transistors, thermally grown silicon oxide has effective trap densities $N_{it} \sim 5 \times 10^{10} cm^{-2}$ and $N_{ot} \sim 5 \times 10^{11} cm^{-2}$ for interface traps and oxide traps, respectively.[27, 28] These concepts are adapted to the graphene/$SiO_2$ interface. Interface traps are populated continuously as the gate voltage is tuned, usually oxide traps are charged only with injection at gate fields above $3 \times 10^{-2} V \cdot nm^{-1}$ before $SiO_2$ breaks down at about $0.27 V \cdot nm^{-1}$.[27] The graphene FET capacitance per unit area is derived from simple capacitor model $C_g = 115\ aF/(\mu m)^2 = 712\ e\cdot(\mu m)^{-2}\cdot V^{-1}$. The number of charges trapped per unit area $N = \dfrac{C_g \cdot \Delta V_{np}}{2e}$ is estimated from the NP shift $\Delta V_{np}$ and an effective capacitance $C_g$. Therefore, the movement of neutrality point can be explained by the effective trapping charges density. Trapped charge density ($N_{10V}$) with $\Delta V_{np} = 0.8V$ in a range of 10V should be $N_{10V} = \dfrac{\Delta V_{np} C_g}{2e} \sim 5.76\times 10^{10}\ cm^{-2}$, trapped charge density ($N_{80V}$) with $\Delta V_{np} = 15V$ in a range of 80V should be $N_{80V} = \dfrac{\Delta V_{np} C_g}{2e} \sim 5.4\times 10^{11}\ cm^{-2}$. The calculated value indicates that the trapped charge density near graphene/SiO$_2$ interface is in the same scale with that near Si/SiO$_2$ interface. Note that the results agree well with the size of charge density nonuniformity ($\delta n = 2 \sim 15\times 10^{11}\ cm^{-2}$ in SLG and BLG) obtained from other experimental methods[4, 6, 29, 30]. The hysteresis clearly demonstrates dynamic duration how the charges inject into trapping centers. Figure 2(d) illustrates the diagram of avalanche injection of holes into interface or bulk oxide traps from the graphene FET channel.

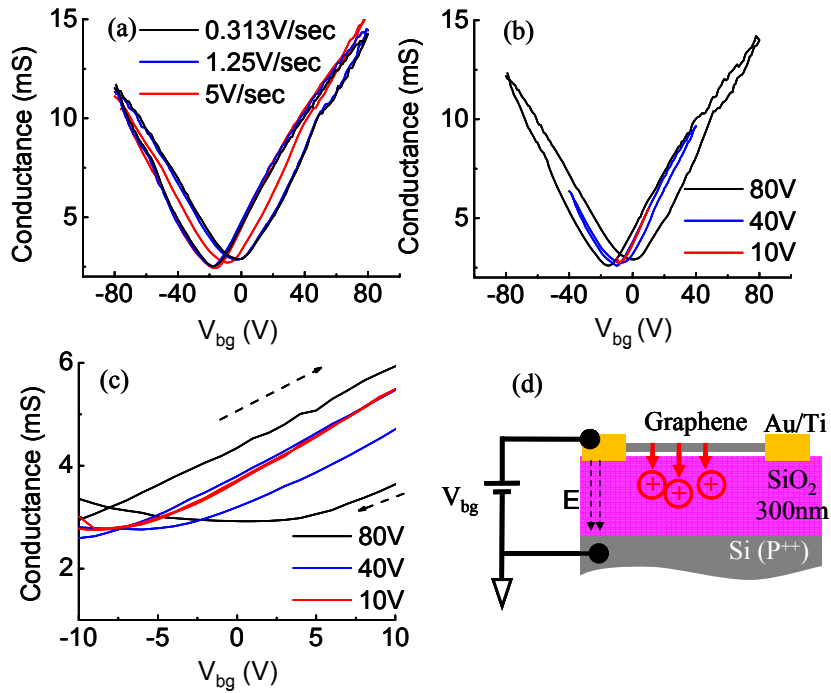

FIGURE 2. a) Conductance hysteresis recorded for the same device as in Fig. 1 under three different V$_{bg}$ sweeping rates in ambient condition; b) Conductance vs gate voltage curves under three different gate voltage range in ambient condition; Device hysteresis increases steadily with increasing voltage range due to avalanche charge injection into charge traps; c) Close up of (b) within the low voltage region; d) Diagram of avalanche injection of holes into interface or bulk oxide traps from the graphene FET channel.

We performed a statistical analysis on many graphene transistors in order to examine the hysteresis strength of graphene flakes ranging from 1 to 10 layers. The graphene flakes are obtained from the same bulk graphite and transferred to the same substrate surface. All the devices were formed together by one round of fabrication processes. The relationship between shift of NP and flake thickness is shown in Figure 3a. The dotted lines serve as a guide for eyes. Note that all samples in Figure 3 were fabricated by method 2 (see Methods). Although the NP shift fluctuates in a large range, it is noticed that the average shift of NP for SLG approaches 12.6 V. However, the NP shift in BLG and FLG starts to deviate from this value; it is about 9.3 V for BLG, 6 V for three-layer stack and saturates at about 5V for thicker layers. The phenomena are due to the charge distribution, which is brought about by the interplay between interlayer hopping and interlayer screening in multilayer graphene. The saturation of NP shift starting from the 4-layer graphene indicates that the carriers induced by external electrical field are mainly located within three near-surface layers. The charge screening length under external field is the thickness of about 2-3 graphene layers near the $SiO_2$ interface. The distribution exists in both the case of A-B (Bernal) stacking [31] and the case of rotationally stacking[32]. We notice that the shift of NP in SLG and BLG fluctuates in a large range. A possible cause could be lateral inhomogeneity, that is, fluctuations in the carriers' concentration, within the thin graphene plane.[4]

The influence of the adsorbates on hysteresis is also interesting. Graphene is always found to be p-type due to unintentional doping of adsorbates (*e.g.*, water, oxygen, organic residue) after exposed to air. The doping states are quite stable even in vacuum. Recently, passing a high current through graphene was demonstrated to heat it locally to high temperature.[33] Current annealing can remove the foreign impurities on the graphene surface, and then improve the quality of graphene. Figure 3 (b) shows the conductance as a function of $V_{bg}$ for sample LF5 before and after current annealing. The annealing and measurements are performed under helium atmosphere at room temperature. The graphene device exhibits p-type before annealing. The DC current is gradually ramped across the device up to 5mA (accordingly about 0.5 mA/μm per layer) and kept for about 10 minutes, and then gradually reduced to zero. After that, the conductance is examined. The procedure is repeated until NP shifts back to zero. As shown in Figure 3 (b), current annealing has been shown to significantly reduce hysteresis. The difference in NP due to hysteresis decreases from 13.6 V ($dV_{gate}/dt$ = 1.25 V/s) to 4.3 V ($dV_{gate}/dt$ = 1.25 V/s). These results suggest that residues may help charge trapping. However, the current annealing treatment could not reduce the hysteresis further. Most likely, impurities permanently trapped in "dead layer" [1] and the $SiO_2$ substrates may be responsible for this lack of further improvement. The possible reason is that the trap sites in $SiO_2$ substrates could not be removed just by this kind of current annealing process. We also notice that current annealing improve carriers mobility from $3200\ cm^2 V^{-1} s^{-1}$ (for holes) and $1400\ cm^2 V^{-1} s^{-1}$ (for electrons) to $8000\ cm^2 V^{-1} s^{-1}$ (for both carriers). The slope of the linear portion of the transfer curve is used to calculate the field effect mobility, $\mu = (1/C_g)\left|\dfrac{d\sigma}{dV_g}\right|$, where

$C_g = \dfrac{\varepsilon \varepsilon_0}{d} = 1.15 \times 10^{-4}\ F/m^2$ is the gate capacitance per unit area ($\varepsilon$ is the dielectric constant of $SiO_2$, $\varepsilon_0$ is the permittivity of free space and thickness of $SiO_2$ d = 300 nm). It is interesting that current heating improves mobility even for samples in contact with substrates. Normally, adsorbates lead to doping, while not causing significant change in carrier mobility.[12] The most possible reason is that the graphene become partly suspended after annealing. Some residues below the graphene migrate and accumulate into bigger particles after annealing. Some parts of the graphene sheet are lifted up and are not in contact with the underlying substrate. An increase of distance between graphene and substrate reduces the influence of the substrate (phonon and trapped charges). The partly free-standing configuration exhibits the intrinsic rippling of graphene[34], which has lower amplitude than that of the surface of the substrate. Both of the causes could increase the mobility of carriers in graphene. In addition, the increase of the spacing between graphene and the substrate may reduce the possibility of charge trapping and create less hysteresis. It is necessary to mention that hysteresis becomes less than 1 V at 4.2 K in helium vapor, even without current annealing. It means the action of trapping is very sensitive to temperature, and it is suppressed at low temperature.

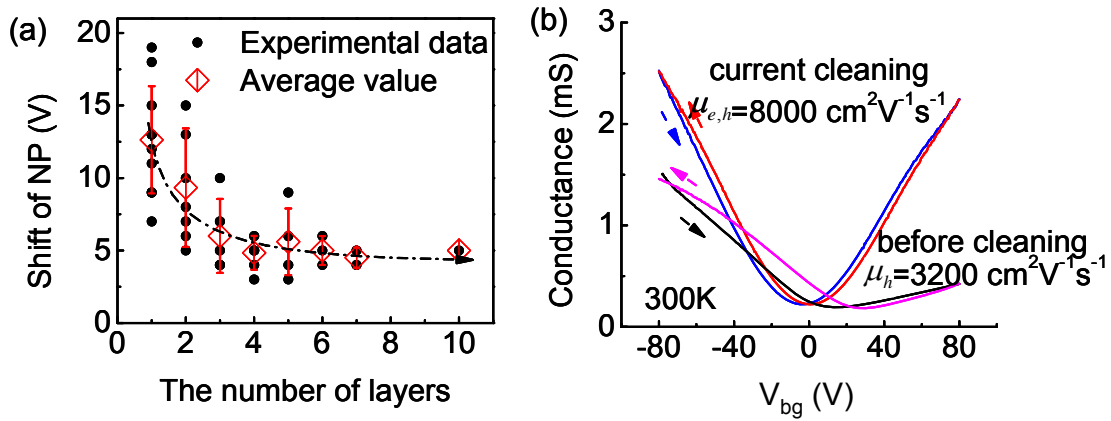

FIGURE 3. a) Shift of the neutrality point as a function of the number of layers. The error bar represents the standard deviation of all the raw data. The dashed line is guide for eyes; b) Two-point conductance as a function of gate voltage in sample LF5 (bilayer) before and after the application of a large current in helium gas atmosphere and at T=300 K.

Graphene FET usually creates a uniform electrical field ($E = \frac{V_g}{d}$, $d$ is the thickness of $SiO_2$) through the gate oxide, while carbon nanotube (CNT) FET generates the radiating electrical field. Electrical field at the interface between CNT and $SiO_2$ follow the relationship ($E = \frac{V_g}{\varepsilon R_t \ln(d/R_t)}$) ($\varepsilon = 3.9$ is the dielectric constant of $SiO_2$, and $R_t$ is the nanotube radius, d = 300 nm is the thickness of $SiO_2$). It can easily reach 1V/nm, greatly larger than the breakdown field of $SiO_2$. The high electrical field near CNT leads to pronounced hysteresis even under a small range of voltage sweeping in CNT transistor.[35-38] Although graphene FETs achieve a comparatively low electrical field, the uniformity of the electrical field near the graphene surface makes graphene a promising candidate in further studies on sensing applications as CNT.

In Figure 4a, charge transfer from/to graphene reduces the charge density in graphene directly and leads to the positive shift of conductance. However, sometimes hysteresis of conductance may go in the opposite direction due to carrier density enhancement in graphene by capacitive gating (Figure 4 (b)). As shown in Figure 4 (b), after applying external electrical field E, ions with inversed charge move toward graphene and dipoles align along the external electrical field. As such, the local electrical field near graphene will be enhanced. The enhancement of local electrical field near graphene helps attract more majority carriers through the metallic contact due to capacitive gating, and then the carrier density in graphene is effectively increased. The negative shift of conductance is finally observed. Figure 4 (c) illustrates this kind of conductance hysteresis in graphene device. This kind of hysteresis can be typically observed in electrolyte-gated graphene device as shown in Figure 4 (d). The sample was fabricated by method 3 (See Methods). The electrolyte is aqueous 1mM KCl solution. The gate voltage was swept continuously within the range of 1V in order to avoid any reaction. Figure 4 (e) compares conductance ~ $V_{tg}$ curves of the electrolyte-gated graphene transistor for different voltage ranges at room temperature. The NP shift almost keeps constant as the range of $V_{tg}$ increases. When the voltage is in a range of 0.5 V, the NP shift due to hysteresis is about 0.25 V. When the range of $V_{tg}$ increases beyond 0.5 V, the hysteresis varies a little. The NP shift goes to about 0.24 V (in the range of 0.8 V) and 0.25 V (in the range of 1 V). The phenomena can be qualitatively explained. For an electrolyte-gated device, the electrode will create an electric field and attract oppositely charged ions from the solution, forming what is know as electrical double layer, the two layers of the charge (surface charge and the layer of counterions) can be approximated very well as a parallel plate capacitors.[39] And the capacitance rises exponentially with the potentials applied onto the electrodes.[40] Similarly, for an electrolyte-gated graphene, when positive voltage is applied to gate (Au), free cations tend to accumulate near the negative electrode (graphene), creating a positive charge layer near graphene. The accumulation is limited by the concentration gradient, which opposes the Coulombic force of the electrical field. The charge layer with inversed charges (called Debye layer) accumulate near the electrolyte and provides a much higher gate capacitance than the commonly used $SiO_2$ back gate, the high capacitance originates from the small distance between Debye layer and electrode (graphene). Therefore, when the gate voltage sweeps, the graphene "remembers" the conductance at the last gate voltage it "saw". The hysteresis of conductance shift left with respect to gate bias, this is because it takes some time for the potential to be distributed in solution. The relaxation time of distribution is not sensitive to sweeping range, because certain biases under similar sweeping rate just accumulate similar concentration of inversed charge at a similar rate. As such, the NP shift in hysteresis almost keeps constant as the range of $V_{top-gate}$ (abbreviated to $V_{tg}$) increases.

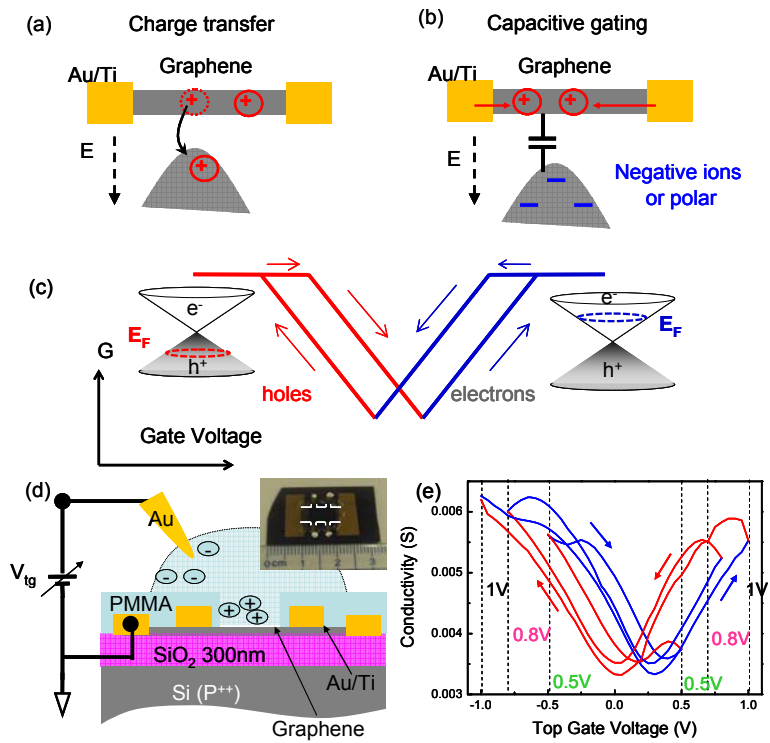

FIGURE 4. The carrier density in graphene is affected by two mechanisms. a) Transferring a charge carrier (hole) from graphene to charge traps causes the right shift of conductance, and vise versa; b) Capacitive gating occurs when the charged ion or polar alters the local electrostatic potential around the graphene, which pulls more opposite charges onto graphene from the contacts; c) Schematics of hysteresis caused by the capacitive gating; d) Schematic diagram of the experimental setup using electrochemical gate. A Hall bar configuration was used for the electrochemical transport measurement at room temperature. (The inset shows an optical image of a monolayer CVD graphene FET device (Sample s-CVD8), and the white dashed line profiles the graphene area); e) G vs $V_{top-gate}$ curves records for the top-gated CVD graphene FET under different sweep ranges (as indicated) in ambient air. The arrows denote the sweeping direction.

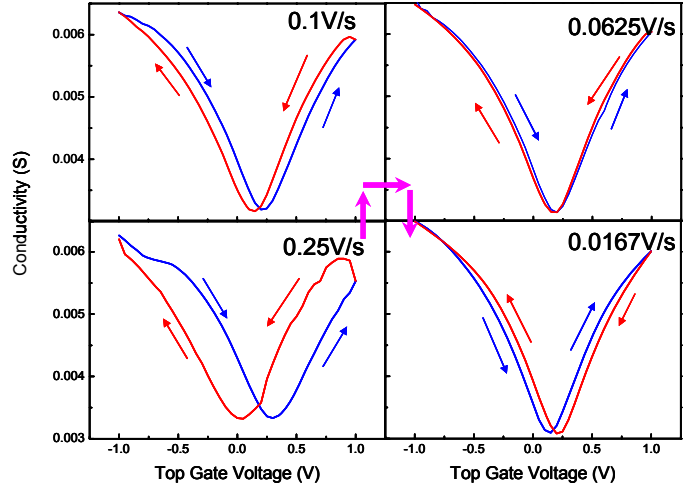

FIGURE 5. Conductance hysteresis recorded under four different $V_{gate}$ sweep rates (0.25V/s, 0.1V/s, 0.0625 V/s, 0.0167V/s) in ambient condition for the sample shown in Figure 4(d). The arrows denote the sweeping direction of gate voltage.

Figure 5 compares conductance ~ $V_{tg}$ curves of the electrolyte-gated graphene FET for different voltage sweeping rates at room temperature. It is found that the hysteresis could be reduced by slowing down the gate sweeping rate. As mentioned previously, the hysteresis of conductance is related to the relaxation time of potential distribution, while the relaxation time of distribution is sensitive to sweeping rate. As such, unbalanced distribution creates larger positive hysteresis. It is also found that the hysteresis finally changed from NP left-shift to NP right-shift. The results indicate that the two mechanisms (charge transfer from/to graphene and capacitive gating effect), generally coexist and compete with each other in graphene transistors. Slow gate sweeping rate could suppress the hysteresis from capacitive gating and enhance the hysteresis from charge trapping. Similarly, the hysteresis from capacitive gating effect could be reduced by increasing the concentration of ions in the solution (as shown in supporting information Figure S4).

We also observed this kind of hysteresis shown in Figure 4(c) in back-gated graphene FET on $SiO_2$ (See Figure S1). Although the NP positively shifts when ramping gate voltage under ambient condition, it shifts toward more negative voltage after the samples were mounted on cryostat in flowing cold helium gas. The results are reproducible at 1.6 K, 4.2 K 50 K and 200 K. The hysteresis phenomena may be caused by the surface dipole moment near graphene such as tape residue[41], atmospheric water,[11, 41] or e-beam resist. The hysteresis in the conductance minima by the sweeping gate at a fixed magnetic field confirms that it is originated by capacitive gating. Similar to an electrolyte-gated device, the carrier density in graphene near the dipoles could be increased by capacitive gating from the dipoles. The dipoles usually sit on graphene with some orientation, and they can flap randomly due to the thermal fluctuation or inhomogeneities without electrical field. After applying an external electrical field, the dipoles will be oriented along the electrical field direction, which causes the enhancement of the local electrical field near dipoles. The carrier density in graphene near dipoles was effectively increased due to capacitive gating from oriented dipoles. In addition, it takes some time for the dipoles to be aligned and settle down under an electrical field. The relaxation time is in the seconds scale. In this way, the hysteresis of conductance shift negatively with respect to gate bias. Similar hysteresis with NP left-shift was also observed in bulk graphene devices from other group[42] at low temperature and graphene transistors on ferroelectric surface[43]. In all graphene devices, the capacitive gating effect faces the keen competition with charge trapping. In most cases, the phenomena are difficult to be observed in graphene lying on $SiO_2$ because the charge transfer may dominate the hysteresis by shifting NP in opposite direction. The negative hysteresis only could be observed when the charge transfer is suppressed at low temperature. It is noted that the hysteresis with conductance minima shifting right was observed in graphene nanoring.[44] The authors attribute the hysteresis to the available trapped states located at the rough edges of graphene nanoribbons. Beside this reason, we believe that the nature of narrow width enhances the local electrical field below the graphene nanoring, analogous to the CNT case discussed earlier. The high electrical field locally ionized the dielectric material below and then caused the charge trapping in graphene.

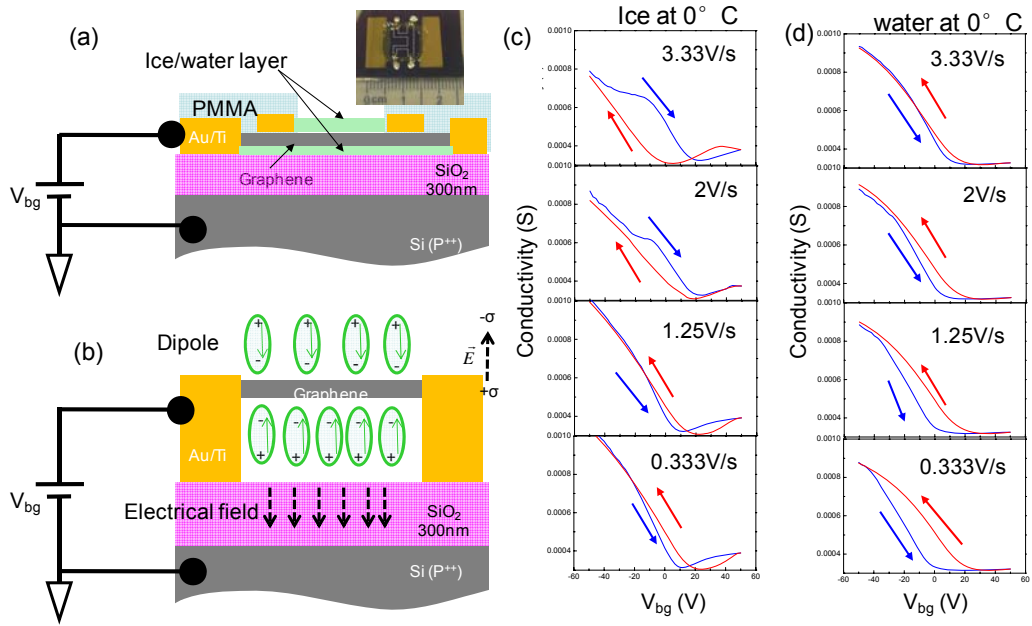

FIGURE 6. Conductance hysteresis in back-gated graphene transistor where the graphene contacts with water/ice. a) Side view of a back gated graphene transistor with four probe configuration. (The inset shows an optical image of a CVD graphene FET device (Sample s-CVD4), and the white dashed lines indicate the graphene area); b) working principle of the CVD graphene FET where water/ice molecules serve as dipoles; pronounced electrical hysteresis curves are observed under four different gate sweeping rates when water c) or ice d) forms on/below graphene at 0°C.

Finally, as water molecules always exist in the surrounding of graphene devices, we would like to investigate the behavior of graphene transistor with water absorbed on graphene flakes. Although a water molecule does not bond strongly with a graphene surface[45], the close contact of graphene with water might be able to modify the electronic properties by serving as trap site or exhibiting a large dipole moment. Here, we investigate the conductance hysteresis of graphene transistors consisting of water film. The back-gated graphene device shown in Figure 6 (a) was fabricated by method 3 (See Methods). PMMA resist was utilized to keep the metal electrode from water film. Water films are formed from atmospheric moisture that has condensed because of the cold substrate. The sample was measured in a refrigerator under atmosphere. Figure 6 (c) and 6 (d) show the typical hysteresis curves of the CVD graphene transistor in the presence of water and ice at 0°C, respectively. The back-gate bias was swept with different rates. It is found that the graphene transistor exhibits obvious negative hysteresis with sweeping rate of 3.33V/s when ice forms on its surface. And the negative hysteresis decays with the decrease of sweeping rate and finally converts to a positive hysteresis. However, the negative hysteresis was not observed in the presence of water even with the gate sweeping rate of 3.33V/s, as shown in Figure 6 (d). Detailed temperature dependence of hysteresis also exhibits the obvious difference between water and ice in different samples (see supporting information Figure S5 and S6). The experimental results clearly show that the ice layer on graphene exhibits stronger average dipole moment than water layer. The results are in good agreement with recent theoretical work[46]. Figure 6 (b) demonstrates that the dipoles of water are oriented along the external electrical field. The manner causes the enhancement of local electrical field near dipoles, which leads to the increase of carrier density in graphene by capacitive gating. As the capacitive gating effect caused by ice dipoles suffers the keen competition with charge trapping, positive hysteresis caused by charge trapping pronounces at low sweeping rate.

## CONCLUSIONS

There are two mechanisms that may cause different hysteresis in graphene during the period of sweeping the gate voltage forward and backward. They are charge transfer and capacitive gating. In both cases, a change of the carrier density in graphene can cause the shift in the conduction with respect to gate voltage. What is more, both of them affect graphene conductance on the seconds time scale. In the first case, when a total charge $\Delta q$ is transferred from\to the graphene, it will cause the positive shift of conductance $\Delta V$ given by $\Delta q = C \Delta V$ where $C$ is the dielectric capacitances. In the second case, capacitive gating increase the carrier density in graphene, and more charges are pulled onto graphene from the metallic contacts. The effect changes the graphene conductance in the same way as gating the transistor through a back gate. The two mechanisms will shift the conductance with respect to gate voltage in the opposite direction. It is also found that an ice layer on/under graphene has a much stronger dipole moment than a water layer. Our experimental results indicate that graphene may offer more opportunities for chemical and biological sensing than a carbon nanotube, due to its huge surface to volume ratio and unique electronic dispersion. In spite of the absence of a band gap in graphene, this hysteretic behavior can be exploited to create nonvolatile memory devices.

## METHODS:

**Device fabrication:**

**Method 1:** Graphene sheets were produced by mechanical exfoliation of bulk graphite using an adhesive tape,[1] then deposited onto Si substrate with a layer of $SiO_2$ 300nm thick. The silicon substrate serving as a global back gate is a highly p-doped (Boron) silicon wafer with a nominal resistivity of less than 0.005 Ohm·cm. After localization of suitable sheets by optical microscopy with respect to alignment marks, metallic leads are patterned using standard electron beam lithography followed by electron beam evaporation of Ti/Au (5nm/80nm thick).

**Method 2:** Graphene sheets were produced by mechanical exfoliation of bulk graphite using an adhesive tape, then deposited onto Si substrate with a layer of $SiO_2$ 300nm thick. The silicon substrate serving as a global back gate is a highly p-doped (Boron) Silicon wafer with a nominal resistivity of less than 0.005 Ohm·cm. Metallic leads are formed by using shield mask (TEM Grid with the bar width of 10μm) to avoid resist residues and followed by electron beam evaporation of Ti/Au (5nm/80nm thick). The spacing between mask and the substrate is about several hundred nanometers which is determined by the thickness of graphite flakes sitting in between the mask and substrates. Due to incomplete contact of the mask and substrate, the electrode metal typically invaded the electrode spacing below the shadow mask by several hundred nanometers. The small values of invading distance are neglectable. The channel length of the devices was maintained at approximately 10μm.

**Method 3:** The large scale CVD-graphene films used here are synthesized on Cu foils by using techniques analogous to the work described in literature[47]. Following that, the Cu is etched away and graphene transferred onto $SiO_2$/Si, with the help of PMMA. Metallic leads were formed by using shield mask made of plastic plate and electron beam evaporation of Ti/Au (5nm/80nm thick). The dimension of the devices goes to centimeter level. The samples were annealed at 400 °C in Ar/$H_2$ (5%) to remove resist residue introduced during the transfer processes.

*Acknowledgement.* The work in NUS is supported under the grant of NRF-CRP project "Graphene Related Materials and Devices" Grant No. R-143-000-360-281. The work in NTU is supported by the Singapore National Research Foundation under NRF RF Award No. NRF-RF2010-07 and MOE Tier 2 MOE2009-T2-1-037.

*Supporting information available:* Description of graphene identification, electrical measurement details, conductance hysteresis in electrolyte gated graphene transistor with respect to gate sweeping rate and solution concentration, conductance hysteresis in back-gated graphene transistor with respect to gate sweeping rate and temperature, leakage current hysteresis versus gate bias in back-gated and electrolyte-gated transistors. This material is available free of charge *via* the Internet at http://pubs.acs.org.

# SUPPORTING INFORMATION

## Hysteresis of Electronic Transport in Graphene Transistors

*Haomin Wang,* [a), *] *Yihong Wu*

*Department of Electrical and Computer Engineering, National University of Singapore, 4 Engineering Drive 3, Singapore 117576*

*Chunxiao Cong, Jingzhi Shang, and Ting Yu* [a)]

*Division of Physics and Applied Physics, School of Physical and Mathematical Sciences, Nanyang Technological University, 21 Nanyang Link, SPMS-PAP-03-17a, Singapore 637371*

### A. Graphene identification and thickness determination:

The 2D peak of the Raman spectrum in monolayer graphene has a full width at half maximum of about 30cm$^{-1}$ which distinguishes a monolayer from few-layer graphene. The thicknesses of the graphene flakes were determined by contrast and Raman spectroscopies.[1, 2]

### B. Electrical measurement of graphene transistors:

The electrical transport properties of the devices are measured under a low driving current (10nA-1μA) by using low-frequency lock-in technique. Beside the noise suppression, another merit of the AC measurement is to avoid the charge distribution for FET channel under different later bias $V_{ds}$ and gate bias $V_{gs}$ in DC measurement.[3] The variation of the charge distribution may shift the NP in DC measurement. The gate voltage is applied by using Keithley 2400 or Keithley 2612A sourcemeter. Some samples were cooled in liquid $^4$He flow of a variable temperature.

a) Electronic mail: haomin.wang@gmail.com; yuting@ntu.edu.sg
∗ Present address: Division of Physics and Applied Physics, SPMS, Nanyang Technological University, Singapore 637371

## C. Electrical conductance hysteresis in graphene transistors:

The conductance bottom corresponds to the position of the neutral point of the graphene channel. The shape of the transfer curve varies depending on the type of electrode metal,[4] invasive effects,[5] substrate,[6] corrugation,[7] device shape,[8] and absorption of surface[9]. However, the species are tangled together and difficult to parameterize separately. As the parts are out of scope of this work, we concentrate on the position of the neutral point rather than the variation in the width of the transfer curve.

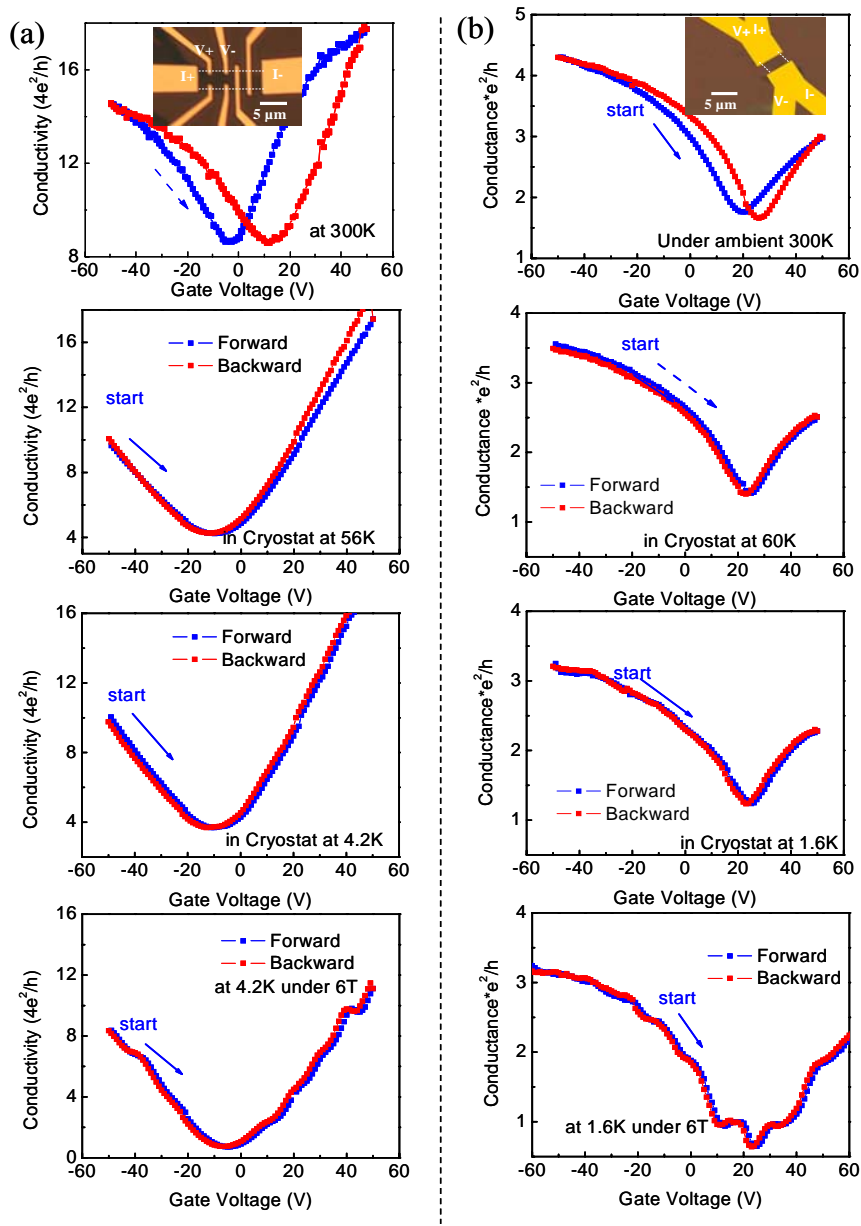

Figure S1. Hysteresis observed in graphene transistors in helium vapor from 300K down to 1.6K with/without perpendicular magnetic fields. (a) 4-layer graphene device

(Bs4q3p8) and (b) monolayer graphene device (Bs5q1p14) are representatives. (The arrows denote the sweeping direction. the insets are optical images of the corresponding devices, and graphene is profiled between dashed lines). Gate sweeping rate is 1.25 V/s.

It is observed that NP positively shifts in back-gated graphene FET on $SiO_2$ when ramping gate voltage at room temperature. However, NP shifts toward more negative voltage after the samples were mounted on cryostat in flowing cold helium gas.

### D. Prototype conductance-switching memory devices

Figure S2 shows the reproducible operation of the prototype conductance-switching memory devices based on graphene channels. Figure S2 (a) shows a schematic diagram illustrating the operating principle of memory devices. When a large negative gate bias is applied, holes tunnel from the graphene to charge trap sites below graphene. After removing the applied gate bias, the remaining positive charges in the NPs repel holes, while attracting electrons in the channels. Here the state of the memory device is written and erased with the gate voltage pulse of 80V and -80V, respectively. The state is then read at $V_{back-gate}=0V$ where the large conductance difference in the hysteresis curve could be obtained.

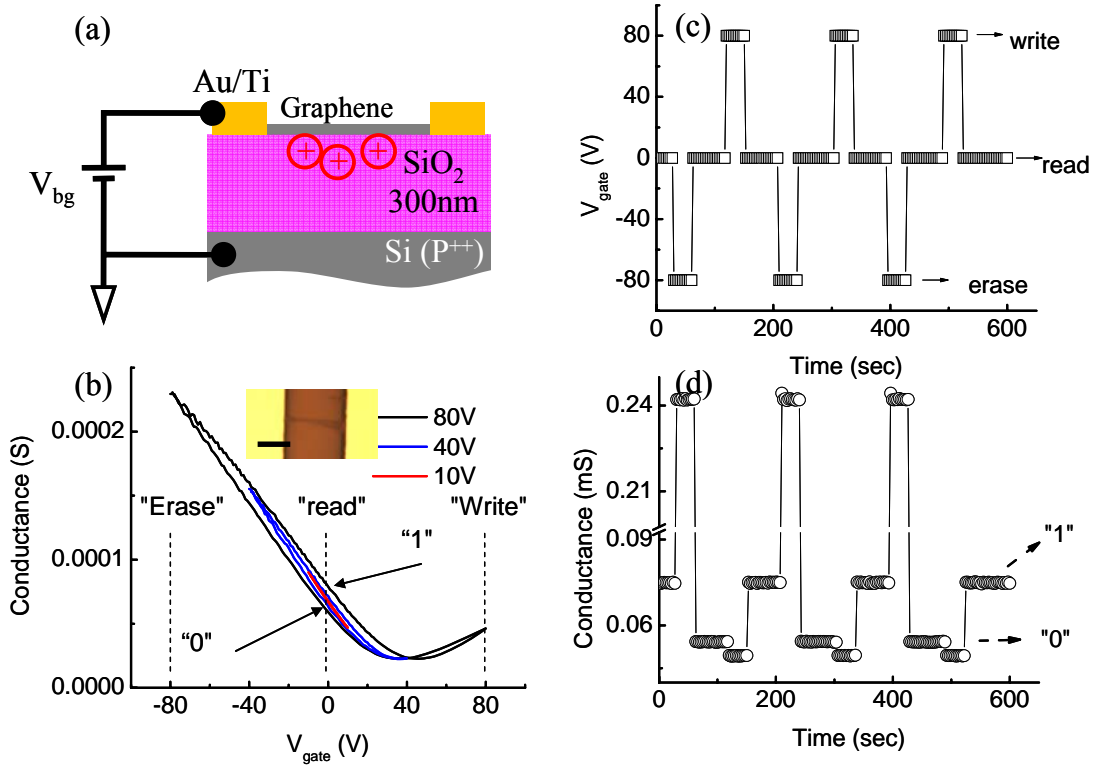

Figure S2. (a) Diagram of holes injection into interface or bulk oxide traps from the graphene FET channel. (b) Conductance as a function of gate voltage at room temperature in sample LF17 within different sweeping range. The inset shows optical microscope topography of the graphene device used in this study, the graphene is the thin and nearly horizontal gray band between the electrodes. Scale bar: 5µm; (c) and (d) three read/write cycles of the graphene FET at room temperature. The lower panel shows the conductance under an AC current of 100nA, while the upper panel shows the gate voltage. The time axis is the same for both graphs. The memory state was read at 0V and written/erased with pulse of ±80V. Sweeping rate=1.25 V/s.

**E. Optical image and Raman spectroscopy of CVD graphene**

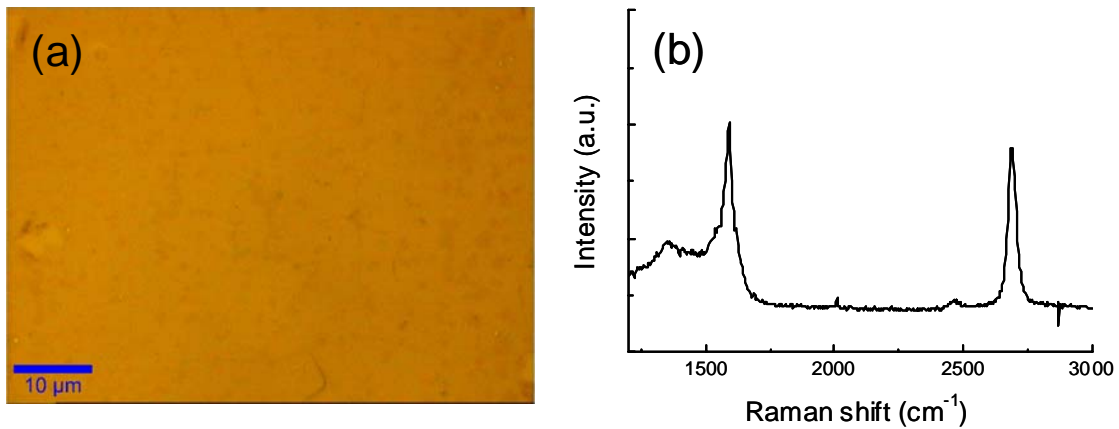

Figure S3. Optical image and Raman spectrum of the CVD graphene.

The Si wafers capped with 300nm $SiO_2$ are ideal substrates for naked eyes to identify graphene layers. The uniform contrast of the optical image indicated the uniformity of graphene thickness. Some small cracks and ripples could be found in graphene film. The cracks were possible formed during the transfer process. A symmetric 2D band centered at 2686cm$^{-1}$ with a full width of half maximum of about 34cm$^{-1}$ determines that the graphene film is monolayer. The weak D band scattering was observed in the Raman spectrum.

**F. Hysteresis in the electrolyte-gated CVD graphene FET with different ion concentration in the solution.**

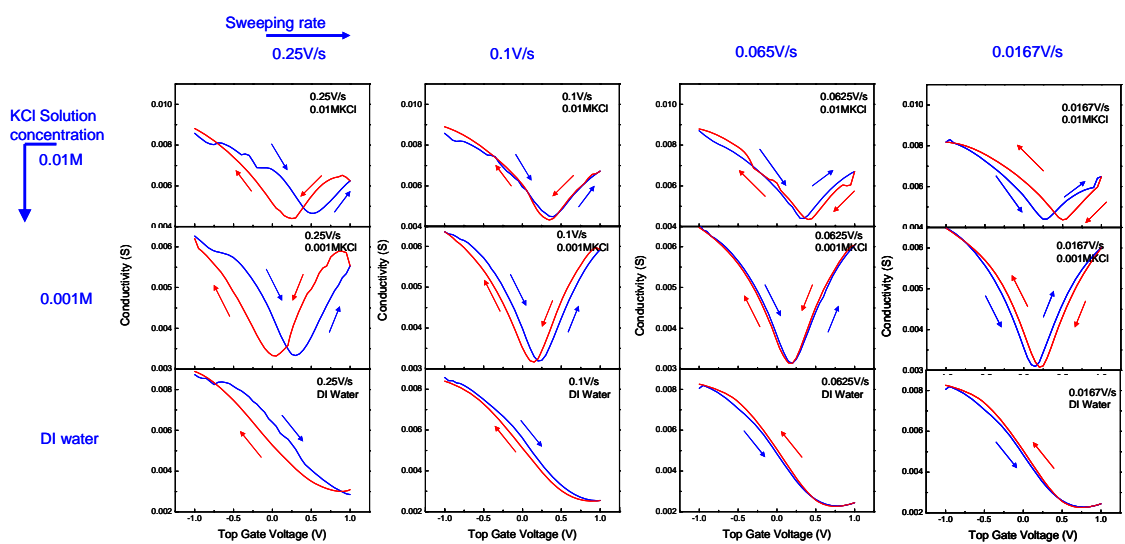

Figure S4. Conductance vs $V_{top\text{-}gate}$ curves records for the electrolyte-gated CVD graphene FET (s-CVD8) under different sweep ranges (0.25V/s, 0.1V/s, 0.0625 V/s, 0.0167V/s) in DI water, or aqueous KCl solution (1mM, 10mM) at room temperature. The arrows denote the sweeping direction.

## G. Hysteresis in back-gated CVD graphene FET with water/ice on/below graphene

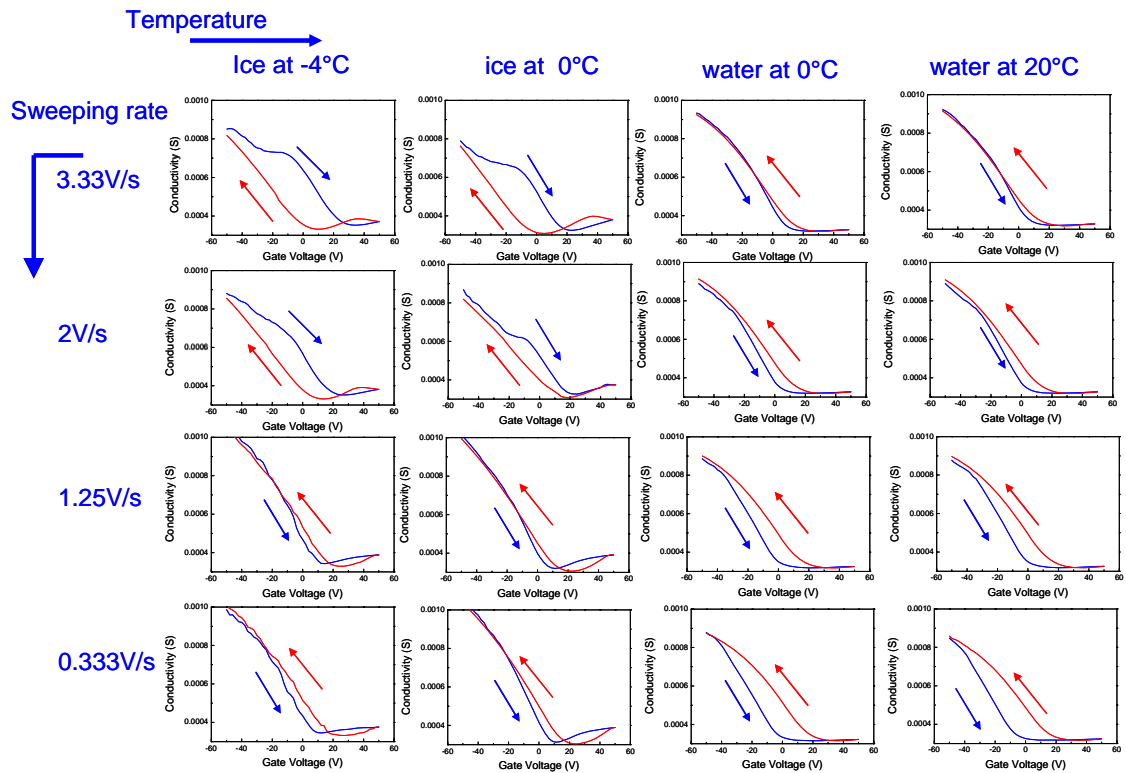

Figure S5. A CVD graphene FET (Sample s-CVD 4) where water/ice molecules serve as dipoles; pronounced electrical hysteresis curves are observed under four different gate sweeping rates when temperature reach (a) -4 °C, 0 °C ( (b) water or (c) ice on/below graphene), d) 20 °C. The arrows denote the sweeping direction.

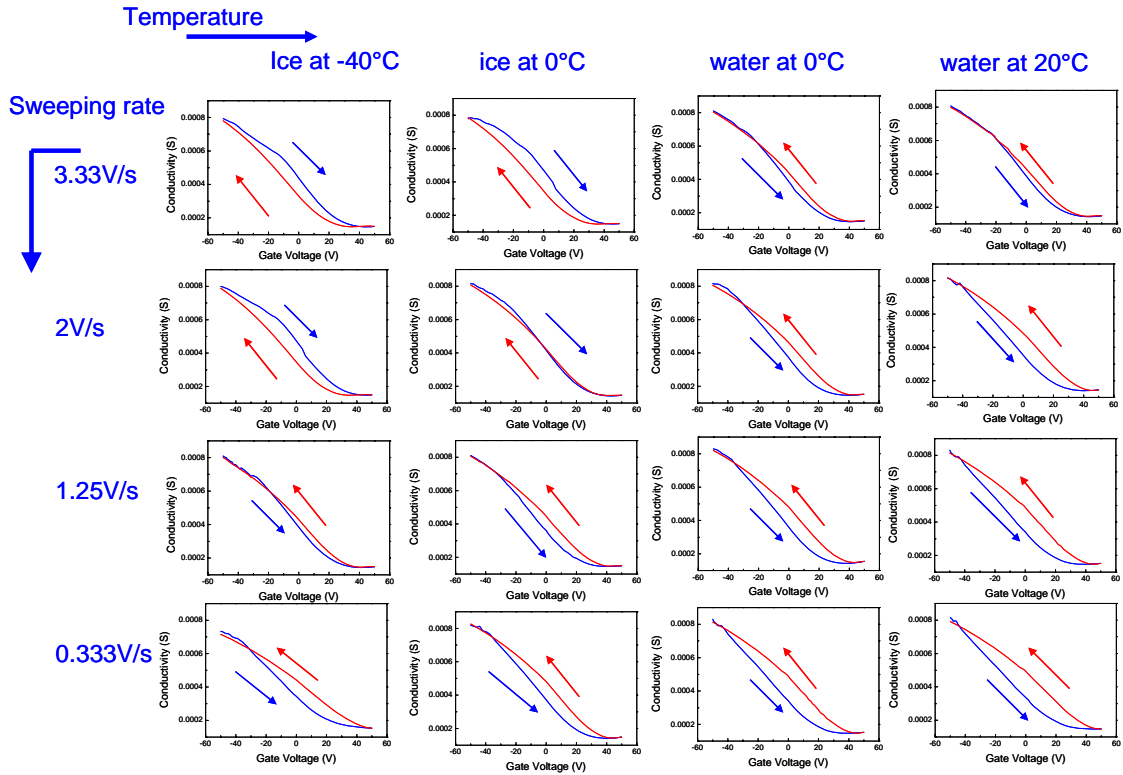

Figure S6. Another CVD graphene FET (Sample s-CVD 11) where water/ice molecules serve as dipoles; pronounced electrical hysteresis curves are observed under four different gate sweeping rates when temperature reach (a) -4 °C, 0 °C ( (b) water or (c) ice on/below graphene), d) 20 °C. The arrows denote the sweeping direction.

In comparison with transistors from exfoliated graphene, transistors from CVD graphene show more obvious positive hysteresis. There are several possible reasons: 1) CVD graphene includes more defects (such as, point defects, grain boundary, edges, ripples and cracks, etc) than the exfoliated one from bulk graphite. The defects could great enhance the local electrical field[10] near them, and finally cause more charge transfer. 2) The residual etchant for Cu foils may serve as charge trap site in CVD graphene.

## H. Gate leakage current

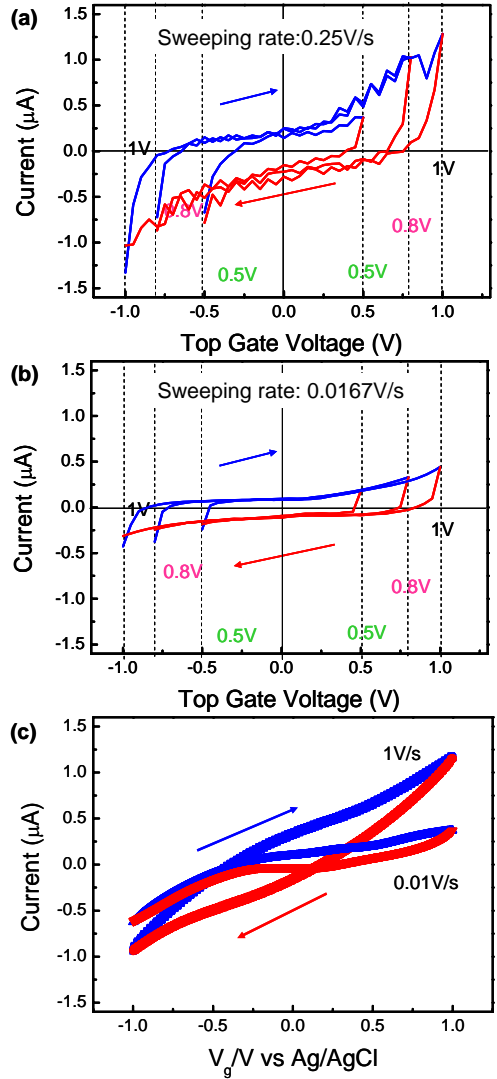

Figure S7. Leakage current ~ $V_{tg}$ curves of the electrolyte-gated graphene FET (Sample s-CVD8) under gate sweeping rate (a) 0.25V/s and (b) 0.0167V/s at room temperature. The data is acquired by sourcemeter Keithley 2400; (c) Leakage hysteresis curves which are obtained by electrochemical workstation CHI760C under two different gate sweeping rates to verify the previous results. The arrows denote the sweeping direction.

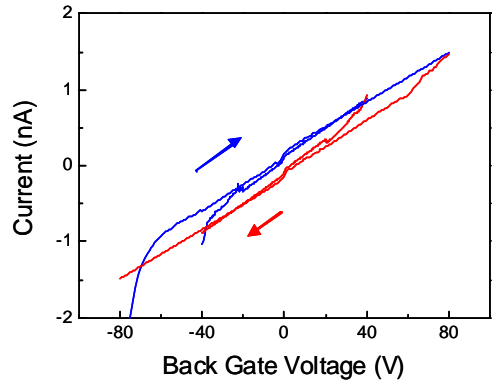

Figure S8. Leakage current ~ $V_{bg}$ curve of the back-gated graphene FET (Sample (bs4q3p7)) obtained by Keithley 2612A under gate sweeping rate of 1.25 V/s. The arrows denote the sweeping direction.